\RequirePackage{fix-cm}
\RequirePackage{amsmath, bbm}
\documentclass[epj,nopacs]{svjour}
\pdfoutput=1
\usepackage{mathtools}
\usepackage{multirow}
\usepackage{axodraw2}
\usepackage{amsmath}
\usepackage{breqn}
\usepackage{url}
\usepackage{bm}
\usepackage{slashed}
\usepackage{scalerel,amssymb}
\usepackage{xspace}
\usepackage{extradefs}

\usepackage{subfig}
\usepackage{tikz}

\newcommand{\swatch}[1]{\tikz[baseline=-0.6ex]\node[fill=#1,shape=rectangle,draw=black,thick,minimum width=5mm,rounded corners=0.5pt](){};}
\newcommand{\met}{\ensuremath{E_T^{\rm miss}}}

\definecolor{darkolivegreen}{HTML}{556B2F}
\definecolor{seagreen}{HTML}{2E8B57}
\definecolor{magenta}{HTML}{FF00FF}
\definecolor{steelblue}{HTML}{4682B4}
\definecolor{mediumseagreen}{HTML}{3CB371}
\definecolor{tomato}{HTML}{FF6347}
\definecolor{salmon}{HTML}{FA8072}
\definecolor{darkorange}{HTML}{FF8C00}
\definecolor{darkorchid}{HTML}{9932CC}

\newcommand{\bsmm}{{$b \rightarrow s \mu^+ \mu^-$}}
\newcommand{\zp}{\ensuremath{Z^\prime}\xspace}
\newcommand{\zz}{\ensuremath{Z^0}\xspace}
\newcommand{\mzp}{\ensuremath{M_{Z^\prime}}\xspace}
\newcommand{\mzz}{\ensuremath{M_{Z^{0}}}\xspace}
\newcommand{\mx}{\ensuremath{M_{X}}\xspace}
\newcommand{\yt}{\ensuremath{Y_{3}}\xspace}
\newcommand{\dyt}{\ensuremath{DY_{3}}\xspace}
\newcommand{\dytp}{\ensuremath{DY_{3}^\prime}\xspace}
\newcommand{\bl}{\ensuremath{{B_3-L_2}}\xspace}
\newcommand{\leg}{The colours indicate the bound giving the dominant sensitivity as in the key below.
    The region above the white solid line is excluded at the 95\% CL and the
    region above the white dashed line is excluded at the 68\% CL.\xspace}

\usepackage{todonotes}
\usepackage{textcomp}

\begin{document}

\title{Large Hadron Collider Constraints on Some Simple $Z^\prime$ Models for $b\rightarrow s \mu^+\mu^-$ Anomalies}  

\author{B.C. Allanach$^1$\and
   J. M. Butterworth$^2$
  \and
  Tyler Corbett$^3$}

\institute{Department of Applied Mathematics and Theoretical Physics, University of Cambridge, Wilberforce Road, Cambridge, 
  CB3 0WA, United Kingdom\\
  B.C.Allanach@damtp.cam.ac.uk
  \and
  Department of Physics \& Astronomy, University College London, Gower St, London,
  WC1E 6BT, United Kingdom\\
  j.butterworth@ucl.ac.uk
  \and
  The Niels Bohr International Academy, Blegdamsvej 17, University of Copenhagen, DK-2100 Copenhagen, Denmark\\
  corbett.t.s@gmail.com} 

\date{}
\abstract{We examine current Large Hadron Collider constraints on some simple
  \zp models that significantly improve on Standard Model
  fits to $b\rightarrow s \mu^+\mu^-$
  transition data. The models that we consider are the `third family baryon
  number minus second family lepton number' (\bl) model and the `third family
  hypercharge' model and variants.
  The constraints are applied on parameter
  regions of each model that fit the $b\rightarrow s \mu^+\mu^-$
  transition data and come from high-mass Drell-Yan di-muons and 
  measurements of Standard Model processes. This latter set of
  observables place particularly strong bounds upon the
  parameter space of the \bl model when the mass of the \zp boson
  is less than 300 GeV.} 

%\emailAdd{B.C.Allanach@damtp.cam.ac.uk}
%\emailAdd{j.butterworth@ucl.ac.uk}
%\emailAdd{corbett.t.s@gmail.com}

\maketitle

%\preprint{MCnet-21-13}

\section{Introduction \label{sec:introduction}}
Certain experimental measurements of particular $B$ hadron decays are
currently in tension
with Standard Model (SM) predictions. The ratios of branching ratios \linebreak
${BR(B \rightarrow K^{(\ast)}\mu^+\mu^-)/BR(B \rightarrow K^{(\ast)} e^+
  e^-)}$~\cite{Aaij:2017vbb,Aaij:2019wad,Aaij:2021vac},
\linebreak${BR(B_s
  \rightarrow
  \mu^+\mu^-)}$~\cite{Aaboud:2018mst,Chatrchyan:2013bka,CMS:2014xfa,Aaij:2017vad,LHCb:2021awg},
angular distributions in\linebreak ${B\rightarrow K^\ast \mu^+ \mu^-}$
decays~\cite{Aaij:2013qta,Aaij:2015oid,Aaboud:2018krd,Sirunyan:2017dhj,Khachatryan:2015isa,Bobeth:2017vxj}
and the branching ratio \linebreak $BR(B
\rightarrow \phi \mu^+ \mu^-)$~\cite{Aaij:2015esa,CDF:2012qwd} are some
examples, which we dub collectively as `\bsmm{} anomalies'. Some of these
observables have 
small theoretical uncertainties in their SM predictions, whereas others have
more sizeable theoretical uncertainties. 
No individual measurement is yet in sufficient tension to claim unambiguous
5$\sigma$ evidence of new physics. However, collectively, the tensions point
to the same conclusion even
when theoretical uncertainties are taken into account: that a beyond the SM (BSM)
contribution to a process connecting a left-handed bottom quark $b_L$, a
left-handed strange quark $s_L$,
a muon $\mu^-$
and an anti-muon $\mu^+$ is preferred (together with the anti-particle copy of
the process).\footnote{Although their tensions with SM predictions are mild, recent experimental determinations of the ratios of
  branching ratios ${BR(B^0 \rightarrow K^{0}_S\mu^+\mu^-)/BR(B^0 \rightarrow
    K^{0}_S e^+ e^-)}$ and ${BR(B^+ \rightarrow K^{(\ast)+}\mu^+\mu^-)/BR(B^+
    \rightarrow K^{(\ast)+} e^+ e^-)}$ bolster this conclusion
  further~\cite{LHCb:2021lvy}.} 
One recent estimate puts the combined global
significance at 3.9 standard deviations~\cite{Lancierini:2021sdf} once the look elsewhere effect
and theoretical uncertainties are taken into account.
Recent fits to single BSM effective field theory operators
broadly agree with each other~\cite{Hurth:2021nsi,Altmannshofer:2021qrr,Alguero:2021anc}: a correction to the
coefficient 
of the effective field theory operator 
$(\overline{b_L} \gamma^\mu s_L) (\bar \mu \gamma_\mu P_X \mu)$ in the
Lagrangian density can greatly
ameliorate the tension between the predictions and the measurements. Here, $P_X$ is a
helicity projection operator. According to the fits, both $P_X=P_L$ (a
coupling to left-handed muons) as well as
$P_X=1$ (i.e.\ a vector-like coupling to muons) work approximately as well as
each other, but $P_X=P_R$ is disfavoured. The fits are typically done in the
approximation that the energy scale (here given
by the bottom meson mass $m_B$) relevant to the measured observables is much
smaller than the scale of new physics producing the operator. 

\begin{figure}
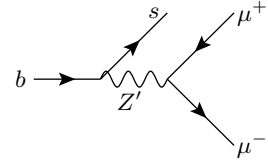

\begin{center}
  \begin{axopicture}(80,55)(-5,-5)
    \Line[arrow](25,25)(50,50)
    \Line[arrow](0,25)(25,25)
    \Line[arrow](75,50)(50,25)
    \Line[arrow](50,25)(75,0)
    \Photon(25,25)(50,25){3}{3}
    \Text(37.5,17)[c]{\zp}
    \Text(-5,25)[c]{$b$}
    \Text(45,50)[c]{$s$}
    \Text(82,50)[c]{$\mu^+$}
    \Text(82,0)[c]{$\mu^-$}    
  \end{axopicture}
\caption{\label{fig:bsmm} A tree-level \zp boson mediated
  contribution to the \bsmm{} transition.}
\end{center}
\end{figure}
A BSM contribution to such effective field theory operators can be generated
by the tree-level exchange of a massive electrically-neutral gauge boson (dubbed
a \zp), as depicted in
Fig.~\ref{fig:bsmm}, if it has family dependent couplings. In particular, to generate 
the effective field theory operator required, it should have a coupling both
to muon/anti-muon fields and to $\overline{b_L} s_L + \overline{s_L} b_L$. 
In the fits of a BSM effective field theory operator to the measurements,
terms of order $m_B^2 / \mzp^2$ are implicitly neglected in the
effective field theory expansion, where \mzp is the mass of the
\zp boson.

\begin{figure}
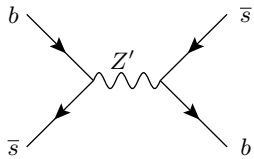

\begin{center}
  \begin{axopicture}(80,55)(-5,-5)
    \Line[arrow](0,50)(25,25)
    \Line[arrow](25,25)(0,0)
    \Line[arrow](75,50)(50,25)
    \Line[arrow](50,25)(75,0)
    \Photon(25,25)(50,25){3}{3}
    \Text(37.5,33)[c]{\zp}
    \Text(-5,0)[c]{$\overline{s}$}
    \Text(-5,50)[c]{$b$}
    \Text(82,50)[c]{$\overline{s}$}
    \Text(82,0)[c]{$b$}    
  \end{axopicture}
\caption{\label{fig:bsbs} Tree-level Feynman diagram of a \zp-mediated process
  which contributes to $B_s-\overline{B_s}$ mixing.}
\end{center}
\end{figure}

In order to explain the \bsmm{} anomalies with such a
\zp, the 
product of the \zp-coupling to $\overline{b_L} s_L + \overline{s_L}
b_L$ and its coupling to $\mu^+\mu^-$ divided by $\mzp^2$ has a range
of values that does not include zero. It is then of interest to ask if such a
\zp boson could be produced directly at high energy proton-proton ($pp$)
colliders and detected so 
as to provide a smoking-gun signal of the model.
The most obvious decay
channel is into $\mu^+\mu^-$: the $\overline{b_L} s_L+ \overline{s_L}
b_L$ coupling is strongly bounded from above by $B_s - \overline{B_s}$ mixing
constraints, where the data and SM
prediction are consistent with each other, but where tree-level \zp contributions
from the diagram in Fig.~\ref{fig:bsbs} are predicted.
It was shown in Refs.~\cite{Allanach:2017bta,Allanach:2018odd} that one
expects a 100 TeV future circular hadron-hadron collider
to be able to cover much of the available parameter space of
generic toy \zp models in the
$\mu^+\mu^-$ channel in parameter regions consistent with the \bsmm{} anomalies. 
Such searches have been
carried out by the Large Hadron Collider (LHC) general purpose experiments
ATLAS and CMS, but so far 
no significant signal has been found. Constraints from a 139 fb$^{-1}$ 13 TeV
$pp$ ATLAS high-mass Drell-Yan di-lepton search were placed upon the toy models in
Ref.~\cite{Allanach:2019mfl}. The parameter spaces of toy models which fit the
\bsmm{} data were only weakly constrained by the
search. However, a more complete model, the Third Family Hypercharge ($Y_3$)
Model~\cite{Allanach:2018lvl}, was found to be more strongly constrained by
the search; in particular, it was found that $\mzp>1.2$ TeV,
calculated in the tree-level limit with an on-shell \zp (and not
including associated production with a jet). 
Here, the
fit to the $b\rightarrow s \mu^+ \mu^-$ 
data was rather crude: dominant tree-level effects from a SM effective field
theory operator were included, and all renormalisation effects were ignored.

In the $Y_3$ model as well as other typical explicit models, massive
\zp gauge bosons originate from 
spontaneously 
broken $U(1)$ gauge symmetries. Several other models with additional $U(1)$ gauge
symmetries\footnote{However, see Ref.~\cite{Chung:2021ekz} for a case where
  the $U(1)$ gauge symmetry comes explicitly embedded within a larger non-abelian
  symmetry.} and  
family dependent charges have been proposed to obtain a \zp with the
correct properties to explain the \bsmm{}
anomalies~\cite{Altmannshofer:2014cfa,Crivellin:2015mga,Crivellin:2015lwa,Crivellin:2015era,Altmannshofer:2015mqa,Sierra:2015fma,Celis:2015ara,Greljo:2015mma,Falkowski:2015zwa,Chiang:2016qov,Boucenna:2016wpr,Boucenna:2016qad,Ko:2017lzd,Alonso:2017bff,Tang:2017gkz,Bhatia:2017tgo,Fuyuto:2017sys,Bian:2017xzg,Alonso:2017uky,Bonilla:2017lsq,King:2018fcg,Duan:2018akc,Kang:2019vng,Calibbi:2019lvs,Altmannshofer:2019xda,Capdevila:2020rrl,Davighi:2020qqa,Allanach:2020kss,Borah:2020swo,Bednyakov:2021fof,Davighi:2021oel,Greljo:2021npi,Wang:2021uqz,Bhatia:2021eco}.
A variant of the $Y_3$ model, the Deformed Third Family Hypercharge (\dyt) was
introduced in Ref.~\cite{Allanach:2019iiy} where limits from the ATLAS
high-mass Drell-Yan di-lepton
search were placed upon it. However, the global fits to the \yt model and \dyt
model have subsequently changed significantly from the inclusion of
precision electroweak observables, which tend to pull the fit more toward the
SM limit~\cite{Allanach:2021kzj}. A new variant (\dytp) was introduced in Ref.~\cite{Allanach:2021kzj}
and a fit of the model to electroweak and \bsmm{} data was performed. The
aforementioned ATLAS high-mass 
Drell-Yan di-lepton search has also
constrained the parameter space of the baryon number minus second family
lepton number (\bl)
model~\cite{Alonso:2017uky,Bonilla:2017lsq}. In the \bl model,
in contrast to the third family hypercharge type 
models, a region of parameter space with $\mzp < 300$ GeV was found to
simultaneously explain
the  \bsmm{} 
anomalies \emph{and}
pass all of the other relevant experimental
constraints~\cite{Bonilla:2017lsq}. Such low values of $M_{Z^\prime}$ are 
unfeasible in third family hypercharge type models because the fit to
electroweak data would become too poor.

Our aim here is to update LHC constraints on the parameter space regions of the \yt,
\dyt, \dytp and \bl models that fit the \bsmm{} anomalies. For
the first three models in this list, the good-fit parameter 
space has changed 
significantly due to the inclusion of electroweak precision observables in the
fit of Ref.~\cite{Allanach:2021kzj} (pertinent new data from LHCb were
included and the theory predictions were improved to include a proper matching to the
SM effective field theory calculation and renormalisation effects, but these 
had a less dramatic effect).
We will check the
constraints coming from measurements of SM-predicted processes
on the \dyt, \dytp and \bl models for the first
time. Notably, we shall 
show that the
$\mzp < 300$ GeV region of the \bl model is strongly
constrained by such measurements. The calculation of high-mass Drell-Yan di-lepton
search constraints upon the 
\dytp model is also new. The calculation of limits in the higher mass
r\'{e}gimes of the models is more accurate than previous determinations in the
literature because it includes associated production of a jet.\footnote{Associated
production of a jet is calculated in the present paper with transverse momentum cut-off at 20~GeV on
the outgoing legs of the two-to-two matrix elements.}

The paper proceeds as follows: the models are defined in
\S\ref{sec:models} along with a characterisation for each model of the parameter region 
that fits the \bsmm{} anomalies.  In \S\ref{sec:const}, we
then go on to 
introduce the various measurements used to constrain the models, which are
calculated by the \CONTUR{2.2.0}
computer program~\cite{Butterworth:2016sqg,Buckley:2021neu}. We present the resulting
constraints in \S\ref{sec:results} before summarising and discussing them
in \S\ref{sec:conc}.

\section{Models \label{sec:models}}
In this section we shall introduce the four models under study in
the present paper.
The chiral
fermion content of each model is that of the SM augmented by three
right-handed neutrinos (SM$+3\nu_R$); these are added in order to obtain
neutrino masses.  
Each model extends the
SM gauge group by an additional $U(1)$ gauge group factor
and each is anomaly-free. In order
to distinguish it from $U(1)_Y$, we shall call the additional $U(1)$ gauge
group $U(1)_X$. 
$U(1)_X$ is broken by the vacuum expectation value $\langle \theta \rangle$ of
a SM-singlet complex scalar $\theta$, the flavon, which has a non-zero $U(1)_X$ charge
$X_\theta$. Then, the $U(1)_X$ gauge boson $X^\mu$ acquires a mass via the
Brout-Englert-Higgs mechanism,
\begin{equation}
  \mx = g_X X_\theta \langle \theta \rangle,
\end{equation}
where $g_X$ is the $U(1)_X$ gauge coupling.
\begin{table}[t]
  \begin{center}
    \begin{tabular}{|c|ccccccc|}\hline
\multirow{2}{*}{$Y_3$} &     $Q_{L_3}^\prime$ & $e^\prime_{R_3}$ &
$u^\prime_{R_3}$ & $d^\prime_{R_3}$ & $L^\prime_{L_2}$ & $H$ &\\
   &   1/6  & -1   &  2/3  & -1/3  & -1/2  & 1/2 &\\ \hline    
\multirow{4}{*}{$DY_3$} &  $Q_{L_3}^\prime$ & $e_{R_3}^\prime$ & $u_{R_3}^\prime$ & $d_{R_3}^\prime$ & $L_{L_2}^\prime$ & $e_{R_2}^\prime$ & $L_{L_3}^\prime$  \\  
   &   1/6 &   -5/3 & 2/3  & -1/2 & 5/6 & 2/3 & -4/3    \\ 
& $e_{R_3}^\prime$ & $H$ &&&&&  \\
 &   -5/3  & 1/2&&&&& \\ \hline
\multirow{4}{*}{$DY_3^\prime$} & $Q_{L_3}^\prime$ & $e_{R_3}^\prime$ & $u_{R_3}^\prime$ & $d_{R_3}^\prime$ & $L_{L_2}^\prime$ & $e_{R_2}^\prime$ & $L_{L_3}^\prime$ \\  
   &   1/6 &   -5/3 & 2/3  & -1/2 & -4/3 & 2/3 & 5/6  \\ 
&$e_3^\prime$ & $H$ &&&&& \\
& -5/3  & 1/2 &&&&& \\ \hline
\multirow{2}{*}{\bl} & $Q_{L_3}^\prime$ & $e_{R_2}^\prime$ & $u_{R_3}^\prime$ &
$d_{R_3}^\prime$ & $L_{L_2}^\prime$ & $\nu_{R_2}^\prime$ & $H$\\
   &     1   & -3    &  1   &   1   &  -3   & -3  & 0 \\
        \hline
    \end{tabular}
  \end{center}
  \caption{\label{tab:charges}$U(1)_X$ charges $X_{\phi^\prime}$ of weak
    eigenbasis fields $\phi^\prime$ in each 
    model.  SM$+3\nu_R$
    fields that are not listed for a model
    have a zero charge. In addition, each model possesses a complex
    scalar flavon field $\theta$ of $U(1)_X$ charge $X_\theta$.}
  \end{table}
The family dependent charges of the other fields are given in
Table~\ref{tab:charges} for the four models under study. 
We use the following notation for the representation of the chiral fermion
fields under 
$(SU(3), SU(2), U(1)_Y)$, where $i \in \{1,2,3\}$ is a family index:
$Q_{L_i}^\prime:=(u_{L_i}^\prime,\ d_{L_i}^\prime)^T \sim ({\bf 3},\ {\bf 2},\ 1/6)$,
$u_{R_i}^\prime \sim ({\bf 3},\ {\bf 1},\ 2/3)$,
$d_{R_i}^\prime \sim ({\bf 3},\ {\bf 1},\ -1/3)$,
$e_{R_i}^\prime \sim ({\bf 1},\ {\bf 1},\ -1)$ and
$L_{L_i}^\prime:=(\nu_{L_i}^\prime,\ e_{L_i}^\prime)^T \sim ({\bf 1},\ {\bf 2},
-1/2)$ and the $L$ and $R$ suffix refers to left and right-handed chiral
fermions, respectively.
The complex scalar Higgs doublet field transforms as $H \sim ({\bf 1},\ {\bf 2},\ 1/2)$. 
In what follows, we denote 3-component column vectors in family space with
bold font, for example ${\bm
  u_L^\prime}:=(u_{L_1}^\prime,\ u_{L_2}^\prime,\ u_{L_3}^\prime)^T$. 

Each of our models then has $X^\mu$-fermion couplings in the Lagrangian
density as dictated by the
charges of the fermions. For each fermion species
\begin{equation}
  F \in \left\{u_L,\ d_L,\ \nu_L,\ e_L,\ u_R,\ d_R,\ \nu_R,\ e_R\right\},
\end{equation}
we have  a Lagrangian density term
\begin{equation}
  {\mathcal L} \supset -g_X \left(
 \overline{{\bm F}^\prime} \slashed{X} \Lambda^{(F)}  {\bm F}^\prime \right),  
\end{equation}
where we have defined the three-by-three
hermitian matrices\linebreak ${\Lambda^{(F)}:=V_F^\dag \xi_F V_F}$ in terms of the 
3 by 3 real diagonal matrices
$\xi^{(F)}:=$diag$(0,\ X_{F_2^\prime},\ X_{F_3^\prime})$. The $V_F$ matrices
are 3 by 3 unitary matrices that transform $F$ from the (primed) weak
eigenbasis to the (unprimed) mass eigenbasis, i.e.\ ${\bm F^\prime}:=V_F {\bm F}$.
The CKM matrix $V$ and the PMNS matrix $U$ are then predicted to be
\begin{equation}
  V = V_{u_L}^\dag V_{d_L}, \qquad
  U = V_{\nu_L}^\dag V_{e_L}, \label{mixings}
  \end{equation}
respectively. 

In order to specify the models further for phenomenological investigation, one
must make some assumptions about the $V_F$ matrices.
They are taken to be consistent with (\ref{mixings}) once empirical inputs are taken
for the central values of the entries of $U$ and $V$.
For simplicity, we take $V_{u_R}=V_{d_R}=V_{e_R}=V_{e_L}=I_3$, the 3 by 3
identity matrix. $V_{\nu_L}=U$ may be fixed by using empirical
inputs for $U$ (however, neutrinos shall play no further role in our
study). We shall require $V_{d_L} \neq I_3$, since we require a coupling
between $d_{L_2}$ and $d_{L_3}$ and the $X^\mu$ boson in order to explain the
\bsmm{} anomalies. 
A `standard
parameterisation'~\cite{ParticleDataGroup:2020ssz} of a 3 by 3 unitary
matrix\footnote{The standard parameterisation parameterises a family of
unitary 3 by 3 matrices that depends only upon one complex phase and three
mixing angles; a more general parameterisation 
would also depend upon five additional complex phases.} was chosen for $V_{d_L}$:
\begin{equation*}
  \left(\begin{array}{ccc}
    c_{12}c_{13} & s_{12}c_{13} & s_{13} e^{-i\delta} \\
    -s_{12}c_{23}-c_{12}s_{23}s_{13}e^{i \delta} & c_{12}c_{23} -
    s_{12}s_{23}s_{13}e^{i\delta} & s_{23}c_{13} \\
    s_{12}s_{23}-c_{12}c_{23}s_{13}e^{i \delta} & -c_{12}s_{23} -
    s_{12}c_{23}s_{13}e^{i\delta} & c_{23}c_{13} \\    
  \end{array} \right),\nonumber
\end{equation*}
where $s_{ij}:=\sin \theta_{ij}$ {and} $c_{ij}:=\cos \theta_{ij}$, {for angles}
$\theta_{ij},\ \delta \in \mathbb{R}$. 
$\theta_{23}$ was allowed to vary, since it is this coupling that controls the
$X^\mu$ coupling to $\overline{b_L} s_L+ \overline{s_L} b_L$.
The assumptions on the angles and phase in $V_{d_L}$ in the fits we use
are different for the third family hypercharge
type models~\cite{Allanach:2021kzj} and the \bl
model~\cite{Allanach:2020kss}, so we consider each in turn. 

\subsection{Third family hypercharge models}
In this subsection, we shall characterise recent global fits 
of the three third family hypercharge type models
under investigation (\yt, \dyt and \dytp) to \bsmm{}
data~\cite{Allanach:2021kzj}.
Aside from the effects of varying $\theta_{23}$, we set $V_{d_L}$ equal to the
CKM matrix. In more detail, 
the other angles and phases in $V_{d_L}$ are set equal to the central values
of the parameters of the CKM matrix inferred by
experimental measurements,  
i.e.\ $s_{12}=0.22650$,
$s_{13}=0.00361$ and ${\delta=1.196}$~\cite{ParticleDataGroup:2020ssz}.

\begin{figure}
  \includegraphics[width=\columnwidth]{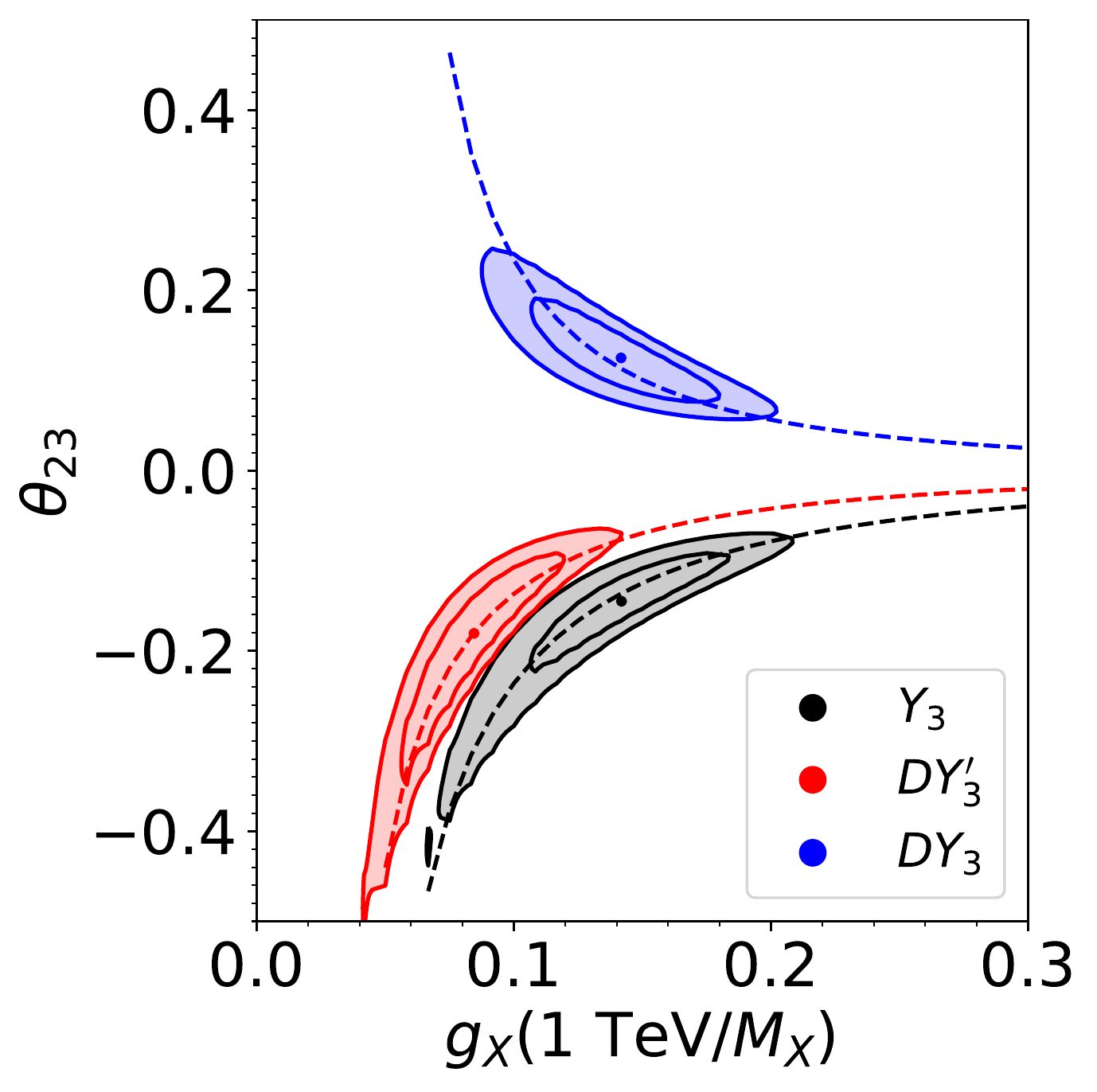}
  \caption{\label{fig:fit} Global fits of third family hypercharge models from
    Ref.~\cite{Allanach:2021kzj} for $\mx=3$ TeV. The
  shaded regions are the $95\%$ CL fit regions for the model, as according to
  the legend. The inner
  contours within each shaded region enclose the 68$\%$ CL region. The
  parameter point
  of best-fit is 
  labelled by a dot in each case.
  To a good approximation, the fits are
  independent of \mx, provided that it is much larger than \mzz and
  provided that $g_X$ is scaled proportional to \mx, as implied by the
  abscissa~\cite{Allanach:2021kzj}. 
  We show our characterisation of the
  good-fit region by the dashed line in each case, detailed in (\protect\ref{t23}) and Table~\protect\ref{tab:param}. } 
\end{figure}

$X_H \neq 0$ in third family
hypercharge type 
models (as Table~\ref{tab:charges} attests) implying that they predict `\zz-\zp
mixing' (i.e.\ mixing between the $X^\mu$ boson, the hypercharge gauge boson
and the electrically neutral $SU(2)_L$ boson) and thus they affect the prediction of
precision electroweak observables.
Ref.~\cite{Allanach:2021kzj} went on to fit the third family hypercharge type
models to 219 electroweak and \bsmm{} data. The
effective field theory calculation of observables in the electroweak sector
implicitly misses relative corrections of ${\mathcal O}(\mzz^2/\mx^2)$ and so is
only valid if $M_{X} \gg \mzz$, the mass of the \zz gauge boson.
After electroweak symmetry breaking, the mass of the gauge boson is corrected
to follow
\begin{equation}
  \mzp^2 = \mx^2[1  + {\mathcal O}(\mzz^2 / \mzp^2)].
  \end{equation}
All three models (\yt, \dyt, \dytp) provided a much better fit than that of
the SM: improving 
$\Delta \chi^2$ by 33--43 units depending upon the model, for
two fitted parameters ($\theta_{23}$ and
$g_X/\mx$). The 95$\%$ confidence level (CL) region
 of each model is shown in Fig.~\ref{fig:fit} as a shaded region, as according
 to the legend.
We wish to reduce the three independent parameters ($g_X, \mx$ and
$\theta_{23}$) down to two in order to display search constraints in terms of
two-dimensional plots. To this end, we characterise each fit in Fig.~\ref{fig:fit}
by a dashed curve. The equation of each
curve is parameterised by
\begin{eqnarray}
    \theta_{23} = \frac{1}{2}\sin^{-1}\left(\frac{a}{x^2 + b x}\right), \label{t23}
\end{eqnarray}
where
\begin{equation}
  x:=g_X (\text{1 TeV}/\mx)
\end{equation}
and $a$ and $b$ were `fit by eye' for each
model. The values taken, along with the domain of good fit, are displayed in
Table~\ref{tab:param} for each model. 
\begin{table}\begin{center}
  \begin{tabular}{|c|cc|c|} \hline
    Model & $a$ & $b$ & $x$ \\ \hline
    \yt & -0.01 & 0.12 &  0.08--0.2 \\
    \dyt & 0.0045 & 0  &  0.1--0.2\\
    \dytp & -0.0045 & 0.067 & 0.04--0.13 \\ \hline
    \bl & -0.0005 & 0 & 0.05--0.62 \\
    \hline    \end{tabular}
  \end{center}
  \caption{\label{tab:param} Parameterisation values and domain of $x$ for the 95$\%$
    CL region for each model, as described by (\protect\ref{t23}). In the \bl model, there is no currently
    calculated upper bound for $x$.}
  \end{table}
It will suit us below to
scan in $x$ and \mx$\approx$\mzp , 
constraining $\theta_{23}$ to satisfy (\ref{t23}) in order to stay within the
region of good fit for the domain of $x$ given in Table~\ref{tab:param}.
We shall refer to this region, where flavour data (and electroweak data for
the third-family hypercharge type models) are within the 95$\%$ CL, as the
`favoured region' of each model.

\subsection{\bl model}
In the \bl model~\cite{Alonso:2017uky,Bonilla:2017lsq}, since $X_H=0$,
there is no
$\zz-\zp$ mixing at tree-level and so $\mzp=\mx$. Electroweak
precision observables then 
follow the SM predictions, to a good 
approximation. We may thus entertain lower values of \mzp, since neither the
theoretical consistency constraint
nor the need to avoid large corrections to SM predictions of  electroweak
observables (both of which imply  $\mx \gg \mzz$
in third
family hypercharge type models)
apply to the \bl model. 
Ref.~\cite{Bonilla:2017lsq} showed that, as well as possessing viable $\mzp>1$
TeV parameter space, 
the model has a region of parameter space within the domain
$20\text{~GeV}<\mx<300\text{~GeV}$
which simultaneously evades other constraints whilst providing much improved fits 
to \bsmm{} data. 

\begin{figure}
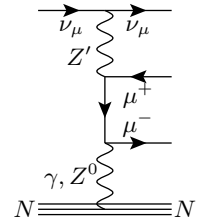

\begin{center}
\begin{axopicture}(55,75)(-5,-2)
  \Line[arrow](0,75)(25,75)
  \Text(12.5,68)[c]{$\nu_\mu$}
  \Line[arrow](25,75)(50,75)
  \Text(37.5,68)[c]{$\nu_\mu$}  
  \Photon(25,75)(25,50){3}{3}
  \Text(17,58)[c]{\zp}
  \Line[arrow](50,50)(25,50)
  \Text(37.5,43)[c]{$\mu^+$}
  \Line[arrow](25,50)(25,25)
  \Text(37.5,32)[c]{$\mu^-$}  
  \Line[arrow](25,25)(50,25)
  \Photon(25,25)(25,0){3}{3}
  \Text(12.5,12.5)[c]{$\gamma,\zz$}
  \Line(0,0)(50,0)
  \Line(0,2)(50,2)
  \Line(0,-2)(50,-2)
  \Text(-5,0)[c]{$N$}
  \Text(55,0)[c]{$N$}
\end{axopicture}
\caption{\label{fig:trident} Tree-level Feynman diagram of a \zp
  contribution to the neutrino trident process. $N$ represents a nucleon.}
\end{center}
\end{figure}
The \bl model was matched to fits of \bsmm{} data
in Ref.~\cite{Allanach:2020kss}. The assumptions on $V_{d_L}$ were equivalent
to taking
$s_{12}=s_{13}=\delta=0$ in (\ref{t23}) and it was found,
after matching to the \bsmm{} fit, that $\theta_{23}$
satisfies (\ref{t23}) with the values shown in Table~\ref{tab:param}.
The upper bound on the domain of $x$ comes from measurements of the trident
process, which roughly agree with SM predictions and so cannot receive large
corrections from the \zp-mediated process shown in
Fig.~\ref{fig:trident}. The lower bound on the domain of $x$ comes from
measurements of $B_s-\overline{B_s}$ mixing, which
bounds the contribution coming from the process in Fig.~\ref{fig:bsbs}. 
We note in passing that the fit we match to in the \bl
model taken is less
sophisticated than the global fits used
for the third family hypercharge type models; the fit to the \bl model was at the
tree-level and did not
include renormalisation group effects. In contrast to the third family
hypercharge type models, the predictions of the \bl model for
electroweak observables are identical to those of the SM and so they were not included
in the fit.

\section{LHC Constraints from {\sc Contur} \label{sec:const}}

The \zp-fermion interactions of the four models that we use have been
encoded into \UFO format~\cite{Degrande:2011ua}\footnote{The \UFO and \FEYNRULES files
  are included in the 
  ancillary information attached to the {\tt arXiv} version of this paper, and will
  be bundled with future releases of
\CONTUR.} by
using \FEYNRULES~\cite{Alloul:2013bka}. Currently, the flavon 
is neglected in the files, however $\zz-\zp$ mixing
effects~\cite{Allanach:2018lvl} 
have been included to leading order in $(\mzz/\mzp)^2$ for the
third-family hypercharge type models. 

The \UFO files are then passed to the \HERWIG{7.2.2}~\cite{Bellm:2015jjp,Bahr:2008pv} event generator,
which calculates
%the implied new tree-level interactions, including
% BCA: new tree-level interactions are calculated in the UFO, I thought
the total width and branching fractions
of the \zp, and generates full final-state simulated $pp$ collision
events\footnote{The main resonant \zp production process cross-section
was checked against \MADGRAPHV{2.6.5} and found to be consistent.}.
These events are passed through the \RIVET{3.1.5}~\cite{Bierlich:2019rhm} library of analyses.
This constitutes a `signal injection' of the putative
BSM contribution to several hundred differential cross-sections measured at the LHC and
stored in \HEPDATA~\cite{Maguire:2017ypu}.
\CONTUR then evaluates whether this BSM contribution would have been visible given the experimental
uncertainties, and if it should have been `seen already', derives an exclusion probability. The
approach is described in more detail in~\cite{Buckley:2021neu}.

\section{Results \label{sec:results}}

\begin{figure}
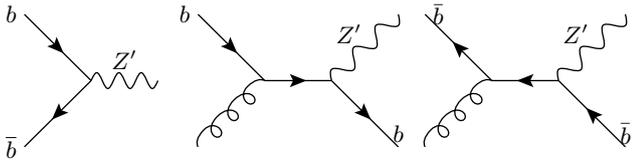

\begin{center}
% Z^\prime diagram
\begin{axopicture}(180,50)(-20,0)
\Line[arrow](-40,50)(-15,25)
\Line[arrow](-15,25)(-40,0)
\Photon(-15,25)(10,25){3}{3}

\Gluon(25,0)(50,25){3}{3}
\Line[arrow](25,50)(50,25)
\Line[arrow](50,25)(75,25)
\Line[arrow](75,25)(100,0)
\Photon(75,25)(100,50){3}{3}
\Text(20,50)[c]{$b$}
\Text(100,5)[c]{$b$}
\Text(81.5,42.5)[c]{$Z^\prime$}

\Gluon(110,0)(135,25){3}{3}
\Line[arrow](135,25)(110,50)
\Line[arrow](160,25)(135,25)
\Line[arrow](185,0)(160,25)
\Photon(160,25)(185,50){3}{3}
\Text(115,50)[c]{$\bar b$}
\Text(185,5)[c]{$\bar b$}
\Text(166.5,42.5)[c]{$Z^\prime$}

\Text(-2.5,33)[c]{$Z^\prime$}
\Text(-45,50)[c]{${b}$}
\Text(-45,0)[c]{$\overline{{b}}$}
\end{axopicture}
\end{center}
\caption{Dominant Feynman diagrams of tree-level inclusive
  $Z^\prime$ production at the
  LHC\@.
  The two right-most diagrams are examples of associated production
  (of a $Z^\prime$ with a $b-$jet or an anti-$b-$jet). 
 \label{fig:Zpprod}}    
\end{figure}
We find that high-mass Drell-Yan LHC searches into a di-muon final state
provide the most constraining bound
at large values of $M_{Z^\prime}$, irrespective of the model. The dominant
partonic $Z^\prime$ 
production processes are shown in Fig.~\ref{fig:Zpprod}, 
where it is 
emphasised that one requires a $b \bar b$ partonic initial state from the LHC
proton pairs $pp$. 
The calculated $Z^\prime$ production cross-section is thus dependent upon which
parton distribution function (PDF) set is used, since the $b(\bar b)$ content of the
proton differs from PDF-set to PDF-set at relatively high values of the ratio of
partonic centre of mass energy to $pp$ centre of mass energy\footnote{We observed
some 20$\%$ differences in the calculated production cross-section when
changing the PDF set used.} For the results discussed below, the default
\HERWIGV{7.2.2} choice of CT14~\cite{Dulat:2015mca} was used.

\begin{figure}[h]
  \centering
  \includegraphics[width=0.4\textwidth]{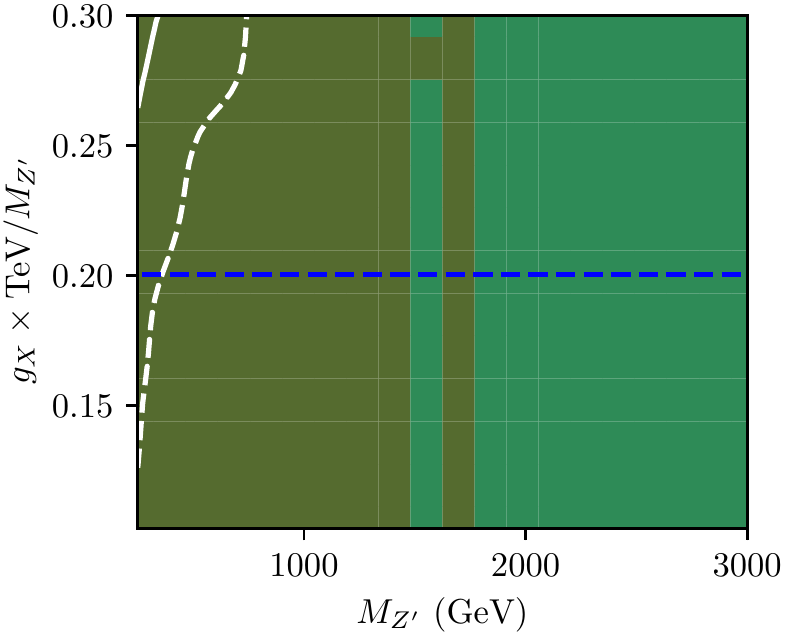}
  \caption{Exclusion in the parameter plane for the \yt model. \leg
    The region favoured by the fits is below the blue dashed line.
    \label{fig:y3}}
    % pool-name legend
    \begin{tabular}{llll}
        \swatch{darkolivegreen}~ATLAS and 
        \swatch{seagreen}~CMS high-mass Drell-Yan $\ell\ell$ 
    \end{tabular}
\end{figure}

The exclusion limit for the \yt model is shown in Fig.~\ref{fig:y3}. The
sensitivity is dominated by the di-muon channel in the high-mass
ATLAS Drell-Yan di-lepton search, with the CMS measurement~\cite{CMS:2018mdl} (which uses
only 3.2 fb$^{-1}$ of integrated luminosity) 
also contributing. The data
barely impinge on the allowed parameter space, with only a small region at low
mass and large coupling disfavoured. Here, we see that the favoured region
$g_X<0.2$ is
not constrained at the 95$\%$ CL in the parameter space considered. 

\begin{figure}[h]
  \centering
  \includegraphics[width=0.4\textwidth]{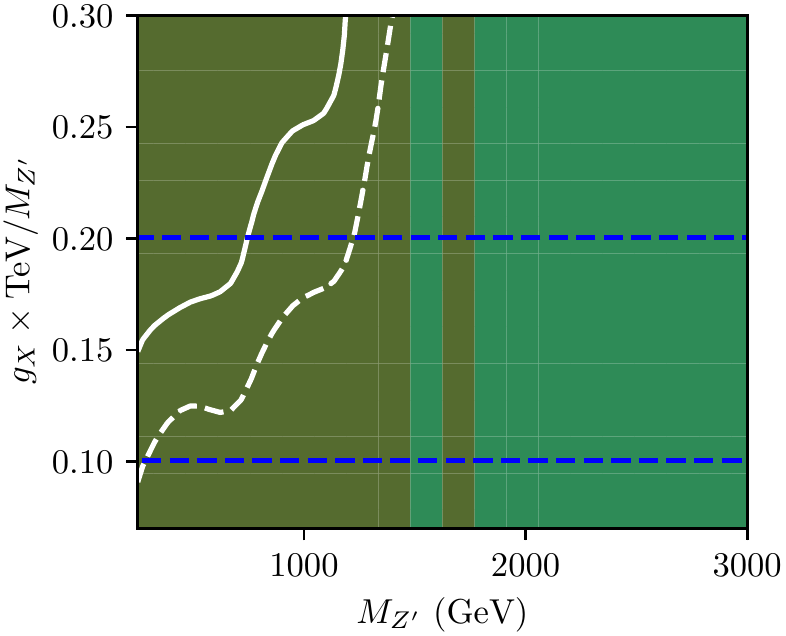}
  \caption{Exclusion in the parameter plane for \dyt model. \leg
    The region favoured by the fits is between the blue dashed lines.
  \label{fig:dy3}}
    % pool-name legend
    \begin{tabular}{llll}
        \swatch{darkolivegreen}~ATLAS and 
        \swatch{seagreen}~CMS high-mass Drell-Yan $\ell\ell$ 
    \end{tabular}
\end{figure}

For the \dyt model, Fig.~\ref{fig:dy3} shows that the sensitivity is somewhat greater,
 extending to $\mzp \approx 1.2$~TeV for the highest values of $g_X \times
 \text{1 TeV}\mzp$ considered, with
 the same two datasets contributing. At low \mzp, the favoured region $0.1 \leq g_X \times
 \text{1 TeV}/M_{Z^\prime}\leq 0.2$ is constrained by
 the ATLAS high-mass Drell-Yan $ll$ search at the 95$\%$ CL. 

\begin{figure}[h]
  \centering
  \includegraphics[width=0.4\textwidth]{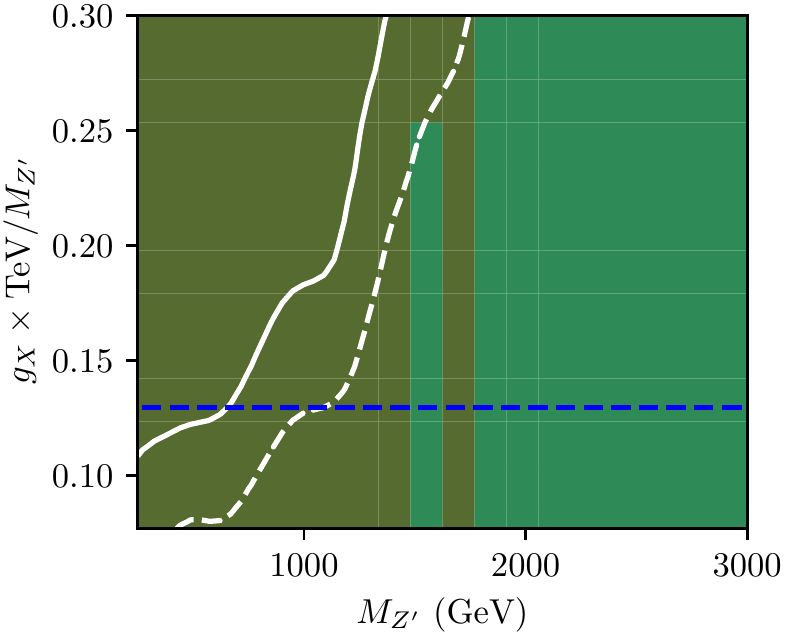}
  \caption{Exclusion in the parameter plane for \dytp model. \leg
    The region favoured by the fits is below the blue dashed line.
  \label{fig:dy3p}}
    % pool-name legend
    \begin{tabular}{llll}
        \swatch{darkolivegreen}~ATLAS and 
        \swatch{seagreen}~CMS high-mass Drell-Yan $\ell\ell$ 
    \end{tabular}
\end{figure}

The exclusion is stronger still for the \dytp model, shown in Fig.~\ref{fig:dy3p}.
At high coupling, masses up to $\mzp \approx 1.3$~TeV are excluded, although in
the region favoured by the fits, only the low \mzp region is impacted.

The comparative strength of the bounds in the three third family hypercharge
models can 
be understood in terms of the size of the absolute additional $U(1)$ charges
of the left-handed and right-handed muons. For a given point in parameter space, the
larger the absolute value of the 
charge, the larger is 
$BR(Z^\prime \rightarrow \mu^+ \mu^-)$ 
and so the bound from high
mass Drell-Yan di-lepton searches is concomitantly stronger\footnote{Strictly
  speaking, 
  $BR(Z^\prime \rightarrow \mu^+ \mu^-)$ also depends upon the
  third family lepton charges (which differ between the three models).
  Taking this into account for the models in question, it is still true that
  $BR(Z^\prime \rightarrow   \mu^+ \mu^-)$ is
  ordered by the comparative absolute value of the muonic charges.}. 
As a careful reading of
Table~\ref{tab:charges} allows, whilst the quark charges are identical for the
three models in question, 
the muon charges are largest for the \dytp
model, next largest for the \dyt model and smallest for the \yt model, 
allowing a rough understanding of the comparative strength of the high mass
Drell-Yan bounds within each.

\begin{figure}[h]
  \centering
  \includegraphics[width=0.4\textwidth]{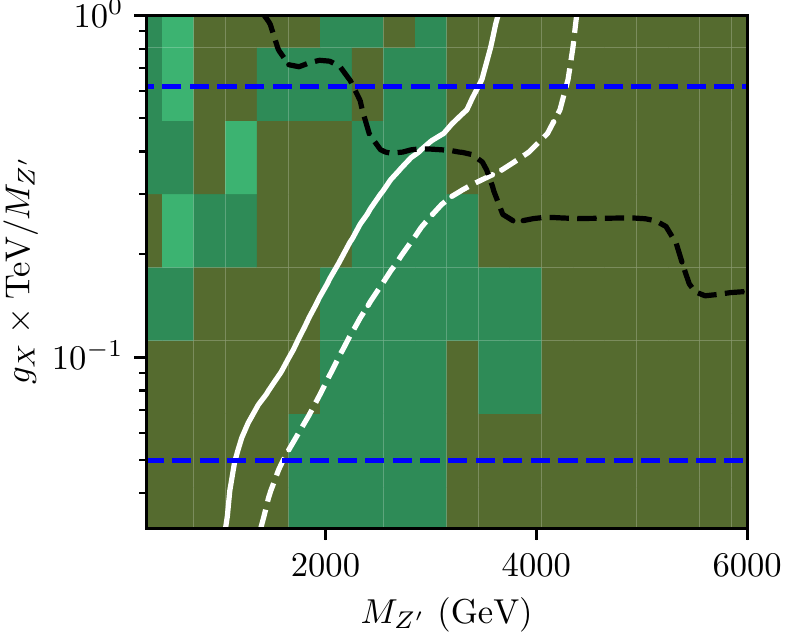}
  \caption{Exclusion in the high mass region of the \bl model. \leg
    In the region above the black dashed line, the width of the \zp is
    more than \mzp/3, and so the perturbative cross-section calculation becomes unreliable.
    The region favoured by the fits is between the blue dashed lines.
    \label{fig:bl1}}
  \begin{tabular}{llll}
        \swatch{darkolivegreen}~ATLAS and 
        \swatch{seagreen}~CMS high-mass Drell-Yan $\ell\ell$ \\
        \swatch{mediumseagreen}~ATLAS $\ell\ell\gamma$ 
    \end{tabular}
\end{figure}

For the \bl model, the range $200 < \mzp < 1000$~GeV is excluded for all allowed couplings, as shown in
Fig.~\ref{fig:bl1}. In this case di-lepton-plus-photon final states~\cite{ATLAS:2019gey,ATLAS:2016qjc,ATLAS:2013way} also contribute.
At high couplings the limit extends up to around 3.7~TeV.
However, as previously mentioned, in this model an open parameter window at low
masses $\mzp<300$~GeV also exists.

\begin{figure}[h]
  \centering
  \includegraphics[width=0.4\textwidth]{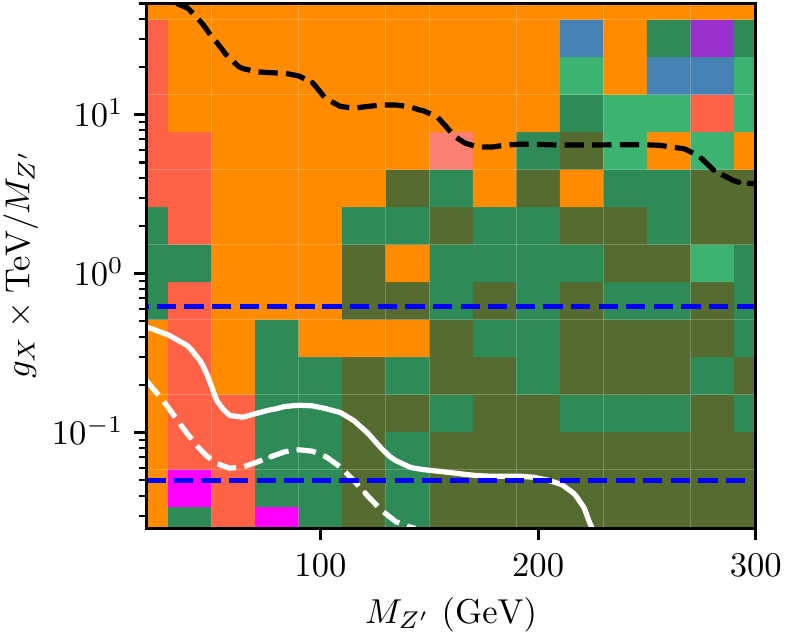}
  \caption{Exclusion in the low mass region of the \bl model. \leg
    In the region above the black dashed line, the width of the \zp is
    more than \mzp/3, and so the perturbative cross-section calculation becomes unreliable.
    The region favoured by the fits is between the blue dashed lines.
    \label{fig:bl2}}
    % pool-name legend
    \begin{tabular}{llll}
        \swatch{steelblue}~CMS $\mu$+\met{}+jet &
        \swatch{darkorchid}~LHCb $\ell$+jet \\
        \swatch{mediumseagreen}~ATLAS $\ell\ell\gamma$ &
        \swatch{tomato}~ATLAS low-mass $\ell\ell$ \\
        \swatch{salmon}~CMS $\ell\ell$+jet &
        \swatch{darkorange}~ATLAS $\mu\mu$+jet \\
        \swatch{darkolivegreen}~ATLAS and &
        \swatch{seagreen}~CMS high-mass $\ell\ell$ \\
        \swatch{magenta}~ATLAS 4$\ell$ 
    \end{tabular}
\end{figure}

In Fig.~\ref{fig:bl2}, we examine this low-mass region. The picture in terms of contributing analyses is more complex.
Di-lepton measurements at the \zz pole, particularly those in Refs.~\cite{ATLAS:2017sag,CMS:2018mdf}, exclude much of the region for\linebreak
${g_X \times \text{1~TeV} / \mzp > 1}$.
Measurements targeted at $W$ bosons decaying leptonically~\cite{LHCb:2016nhs,CMS:2018dxg,ATLAS:2019hxz} play the dominant
role around $\mzp = 300$ GeV for high couplings.
Photon-plus-di-lepton measurements again contribute, especially at higher masses and lower couplings.
Lower mass di-lepton measurements~\cite{ATLAS:2014ape} contribute for $\mzp<\mzz$.
In the region where the sensitivity runs out, the inclusive four-lepton measurement~\cite{ATLAS:2021kog} also contributes, due
to a clean (but very low cross-section) (\zp\zz)-production contribution. With
more integrated luminosity this observable would become more sensitive.

The fact that the cross-section for this model is large enough to significantly distort
the expected distributions, even in the presence of a large SM cross-section,
means that the model
is disfavoured over the majority of the previously open parameter window, leaving only a
small region at low mass and low values of $g_X\times \text{1~TeV}/\mzp$ still
allowed. We note that 
towards the left-hand side of the plot, the accuracy of the previous fit of
the model to flavour data is called into question since unaccounted-for
relative corrections ${\mathcal O}(m_B^2/\mzp^2)$ become sizeable.

For an example point in parameter space for each model,
we display some relevant cross-sections in Table~\ref{tab:xsecs}.

\begin{table}\begin{center}
  \begin{tabular}{|c|c|c|c|c|c|} \hline 
    Model & \mzp    &  $x$  & $\sigma_{\zp\rightarrow\mu^+\mu^-}$ & $\sigma_{\zp+q,g}$ & $\sigma_{\zp+\gamma}$  \\ 
          & (GeV)   &       & Exclusive                       &                  &              \\ \hline 
    \yt   &  540    & 0.2   &  0.7                            & 10               & 0.02         \\ 
    \dyt  &  540    & 0.2   &  3.3                            & 31               & 0.06         \\
    \dytp &  540    & 0.2   &  5.7                            & 26               & 0.07         \\ \hline
    \bl   &   60    & 0.074 &  600         & 750              & 0.4 \\
    \hline    \end{tabular}
  \end{center}
  \caption{\label{tab:xsecs}
    Example leading order matrix element cross-sections calculated by \HERWIG for some parameter points for the four models considered.
    The cross-sections quoted
    are $pp$ cross-sections at 13~TeV centre-of-mass energy, in fb.
    $\sigma_{\zp+q,g}$ is
    the cross-section for associated $Z^\prime$-quark production plus the
    cross-section for associated $Z^\prime$-gluon production, with a minimum
    transverse momentum of 20~GeV for the quark or gluon.
}
\end{table}

\section{Summary and Discussion\label{sec:conc}}

We have calculated LHC bounds upon four models that have been fitted to \bsmm{}
anomalies. Each of the models includes an electrically neutral, massive
$Z^\prime$ gauge boson which has family dependent couplings to SM fermions,
the most important for our discussion being the couplings to $\mu^+\mu^-$ and to $\bar
b s + b \bar s$.\footnote{In principle, a $Z^\prime$ coupling to $\mu^+ \mu^-$ can
change the prediction of the anomalous magnetic moment of the muon $a_\mu$,
which has been measured to be in tension with its SM
prediction~\cite{Abi:2021gix}. However, in order to satisfy other experimental
constraints, simple $Z^\prime$ models such as those deployed in the present
paper are 
forced into a parameter space where the beyond-the-SM contribution to $a_\mu$
from the 
$Z^\prime$
is too small to explain the discrepancy and so further model building
involving additional 
fields and/or additional $Z^\prime$ couplings~\cite{Jegerlehner:2009ry} would be required to
explain the measured value of $a_\mu$.}

Three third family hypercharge type models (the \yt, \dyt and
\dytp models) were recently fit~\cite{Allanach:2021kzj} to \bsmm{}
data, which is in tension with SM predictions. 
The fits included electroweak data, since the third family hypercharge models
alter the SM 
predictions of the electroweak observables.
In the present paper, we have calculated LHC
constraints upon the parameter space of these three models that fit
the \bsmm{} data. The LHC constraints upon the three
third family hypercharge models are
much weaker than those calculated previously in
Refs.~\cite{Allanach:2019mfl,Allanach:2019iiy}. This is due to
the fact both the electroweak data and recent \bsmm{} data pushed the fit
towards the SM limit, 
preferring smaller values of $g_X \times \text{1~TeV}/M_{Z^\prime}$ as
compared to previous preferred values. Our
calculation is at a higher level of precision 
compared to previous estimates, since it includes the effects of associated
production and 
renormalisation of the SM effective field theory in the fit. Neither the \dyt
model nor the \dytp model had been checked previously against measurements of
SM-predicted processes. However, for the parameter regions of interest where
$M_{Z^\prime} > 300$ GeV, these are not as constraining as the high-mass
Drell-Yan di-lepton searches, as Figs.~\ref{fig:y3}-\ref{fig:dy3p} show.
In general even the high-mass Drell-Yan limits
upon the third family hypercharge models
are not very constraining;
one cannot quote a lower bound on \mzp independent of the coupling for any of
the three models. 

In contrast, the \bl model is more tightly constrained by the 
high-mass Drell-Yan di-lepton searches at the
LHC, as Fig.~\ref{fig:bl1} shows: \mzp$>$1 TeV in the favoured region.
The measurements of SM-predicted quantities (calculated here for the first
time) have a large impact on the low
\mzp window of the \bl model, as shown in
Fig.~\ref{fig:bl2}. Differential cross-sections in an ATLAS $\mu\mu+$jet
analysis play a particularly important role at low \mzp. The constraints in
the low \mzp window
are significantly stronger than those calculated previously in
Refs.~\cite{Bonilla:2017lsq,Allanach:2020kss}. 

Generally, measuring $\tau$ leptons is more difficult than measuring muons in LHC
experiments. However, at higher transverse momenta, the hadronic $\tau$
energy resolution is expected to improve~\cite{ATLAS:2017mpa}, since it is
dominated by calorimetery, whereas the muon
resolution degrades, due to the lower curvature in the
trackers~\cite{CMS:2018rym,ATLAS:2020auj}. Higher 
values of \mzp lead to final state particles at higher transverse
momenta on average, so $\tau$ leptons may become relatively more important. 
We note that while no relevant measurements of $\tau$ final states are currently available in \RIVET,
for the \yt model, the branching fraction $\zp \rightarrow \tau^+\tau^-$ is
around\footnote{We quote representative values of the branching fractions in the r\'{e}gime where
  \mzp is much greater than twice the top quark mass. The predicted branching
  fractions we quote here all increase  
  for lower values of \mzp.} 0.3, compared to
$\zp \rightarrow \mu^+\mu^- \approx 0.075$. For \dyt the corresponding branching fractions
are 0.4 and 0.1,
and for \dytp 0.35 and 0.22, suggesting that $\tau$ measurements could make an
important contribution in future, despite the additional experimental
challenges involved. 

For each of the four models that we analyse, an appreciable portion of
parameter space remains where future 
LHC analyses may  search for (and hopefully find) a signal.

\section*{Acknowledgements}
This work has been partially supported by STFC consolidated grants
ST/N000285/1, ST/P000681/1 and \linebreak ST/T000694/1.
This work has received funding from the European Union's 
Horizon 2020 research and innovation programme as part of the 
Marie Skłodowska-Curie Innovative Training Network MCnetITN3 (grant agreement
no. 722104).
BCA thanks other members of the Cambridge Pheno
Working Group for helpful interactions. We especially thank H Banks for helpful
high-mass Drell-Yan di-lepton cross-section and limit comparisons in the $Y_3$ model.

%% For arxiv version, use the next line
\bibliographystyle{JHEP-2}
%% For EPJC version, use the next line
%\bibliographystyle{spphys}

\bibliography{currentLim,contur-anas}

\providecommand{\href}[2]{#2}\begingroup\raggedright\begin{thebibliography}{10}

\bibitem{Aaij:2017vbb}
{\bf LHCb} Collaboration, R.~Aaij {\em et.~al.}, {\it {Test of lepton
  universality with $B^{0} \rightarrow K^{*0}\ell^{+}\ell^{-}$ decays}},  {\em
  JHEP} {\bf 08} (2017) 055 [\href{http://arXiv.org/abs/1705.05802}{{\tt
  1705.05802}}].

\bibitem{Aaij:2019wad}
{\bf LHCb} Collaboration, R.~Aaij {\em et.~al.}, {\it {Search for
  lepton-universality violation in $B^+\to K^+\ell^+\ell^-$ decays}},  {\em
  Phys. Rev. Lett.} {\bf 122} (2019), no.~19 191801
  [\href{http://arXiv.org/abs/1903.09252}{{\tt 1903.09252}}].

\bibitem{Aaij:2021vac}
{\bf LHCb} Collaboration, R.~Aaij {\em et.~al.}, {\it {Test of lepton
  universality in beauty-quark decays}},
  \href{http://arXiv.org/abs/2103.11769}{{\tt 2103.11769}}.

\bibitem{Aaboud:2018mst}
{\bf ATLAS} Collaboration, M.~Aaboud {\em et.~al.}, {\it {Study of the rare
  decays of $B^0_s$ and $B^0$ mesons into muon pairs using data collected
  during 2015 and 2016 with the ATLAS detector}},  {\em JHEP} {\bf 04} (2019)
  098 [\href{http://arXiv.org/abs/1812.03017}{{\tt 1812.03017}}].

\bibitem{Chatrchyan:2013bka}
{\bf CMS} Collaboration, S.~Chatrchyan {\em et.~al.}, {\it {Measurement of the
  $B^0_s \to \mu^+ \mu^-$ Branching Fraction and Search for $B^0 \to \mu^+
  \mu^-$ with the CMS Experiment}},  {\em Phys. Rev. Lett.} {\bf 111} (2013)
  101804 [\href{http://arXiv.org/abs/1307.5025}{{\tt 1307.5025}}].

\bibitem{CMS:2014xfa}
{\bf CMS, LHCb} Collaboration, V.~Khachatryan {\em et.~al.}, {\it {Observation
  of the rare $B^0_s\to\mu^+\mu^-$ decay from the combined analysis of CMS and
  LHCb data}},  {\em Nature} {\bf 522} (2015) 68--72
  [\href{http://arXiv.org/abs/1411.4413}{{\tt 1411.4413}}].

\bibitem{Aaij:2017vad}
{\bf LHCb} Collaboration, R.~Aaij {\em et.~al.}, {\it {Measurement of the
  $B^0_s\to\mu^+\mu^-$ branching fraction and effective lifetime and search for
  $B^0\to\mu^+\mu^-$ decays}},  {\em Phys. Rev. Lett.} {\bf 118} (2017), no.~19
  191801 [\href{http://arXiv.org/abs/1703.05747}{{\tt 1703.05747}}].

\bibitem{LHCb:2021awg}
{\bf LHCb} Collaboration, R.~Aaij {\em et.~al.}, {\it {Measurement of the
  $B^0_s\to\mu^+\mu^-$ decay properties and search for the $B^0\to\mu^+\mu^-$
  and $B^0_s\to\mu^+\mu^-\gamma$ decays}},
  \href{http://arXiv.org/abs/2108.09283}{{\tt 2108.09283}}.

\bibitem{Aaij:2013qta}
{\bf LHCb} Collaboration, R.~Aaij {\em et.~al.}, {\it {Measurement of
  Form-Factor-Independent Observables in the Decay $B^{0} \to K^{*0} \mu^+
  \mu^-$}},  {\em Phys. Rev. Lett.} {\bf 111} (2013) 191801
  [\href{http://arXiv.org/abs/1308.1707}{{\tt 1308.1707}}].

\bibitem{Aaij:2015oid}
{\bf LHCb} Collaboration, R.~Aaij {\em et.~al.}, {\it {Angular analysis of the
  $B^{0} \to K^{*0} \mu^{+} \mu^{-}$ decay using 3 fb$^{-1}$ of integrated
  luminosity}},  {\em JHEP} {\bf 02} (2016) 104
  [\href{http://arXiv.org/abs/1512.04442}{{\tt 1512.04442}}].

\bibitem{Aaboud:2018krd}
{\bf ATLAS} Collaboration, M.~Aaboud {\em et.~al.}, {\it {Angular analysis of
  $B^0_d \rightarrow K^{*}\mu^+\mu^-$ decays in $pp$ collisions at $\sqrt{s}=
  8$ TeV with the ATLAS detector}},  {\em JHEP} {\bf 10} (2018) 047
  [\href{http://arXiv.org/abs/1805.04000}{{\tt 1805.04000}}].

\bibitem{Sirunyan:2017dhj}
{\bf CMS} Collaboration, A.~M. Sirunyan {\em et.~al.}, {\it {Measurement of
  angular parameters from the decay $\mathrm{B}^0 \to \mathrm{K}^{*0} \mu^+
  \mu^-$ in proton-proton collisions at $\sqrt{s} = $ 8 TeV}},  {\em Phys.
  Lett. B} {\bf 781} (2018) 517--541
  [\href{http://arXiv.org/abs/1710.02846}{{\tt 1710.02846}}].

\bibitem{Khachatryan:2015isa}
{\bf CMS} Collaboration, V.~Khachatryan {\em et.~al.}, {\it {Angular analysis
  of the decay $B^0 \to K^{*0} \mu^+ \mu^-$ from pp collisions at $\sqrt s = 8$
  TeV}},  {\em Phys. Lett. B} {\bf 753} (2016) 424--448
  [\href{http://arXiv.org/abs/1507.08126}{{\tt 1507.08126}}].

\bibitem{Bobeth:2017vxj}
C.~Bobeth, M.~Chrzaszcz, D.~van Dyk and J.~Virto, {\it {Long-distance effects
  in $B\rightarrow K^*\ell \ell $ from analyticity}},  {\em Eur. Phys. J. C}
  {\bf 78} (2018), no.~6 451 [\href{http://arXiv.org/abs/1707.07305}{{\tt
  1707.07305}}].

\bibitem{Aaij:2015esa}
{\bf LHCb} Collaboration, R.~Aaij {\em et.~al.}, {\it {Angular analysis and
  differential branching fraction of the decay $B^0_s\to\phi\mu^+\mu^-$}},
  {\em JHEP} {\bf 09} (2015) 179 [\href{http://arXiv.org/abs/1506.08777}{{\tt
  1506.08777}}].

\bibitem{CDF:2012qwd}
{CDF collaboration}, {\it {Precise Measurements of Exclusive b
  \textrightarrow{} s\textmu{}+\textmu{} \ensuremath{-} Decay Amplitudes Using
  the Full CDF Data Set}},  {\em CDF-NOTE-10894} (6, 2012).

\bibitem{LHCb:2021lvy}
{\bf LHCb} Collaboration, R.~Aaij {\em et.~al.}, {\it {Tests of lepton
  universality using $B^0\to K^0_S \ell^+ \ell^-$ and $B^+\to K^{*+} \ell^+
  \ell^-$ decays}},  \href{http://arXiv.org/abs/2110.09501}{{\tt 2110.09501}}.

\bibitem{Lancierini:2021sdf}
D.~Lancierini, G.~Isidori, P.~Owen and N.~Serra, {\it {On the significance of
  new physics in $b\to s\ell^+\ell^-$ decays}},
  \href{http://arXiv.org/abs/2104.05631}{{\tt 2104.05631}}.

\bibitem{Hurth:2021nsi}
T.~Hurth, F.~Mahmoudi, D.~M. Santos and S.~Neshatpour, {\it {More Indications
  for Lepton Nonuniversality in $b \to s \ell^+ \ell^-$}},
  \href{http://arXiv.org/abs/2104.10058}{{\tt 2104.10058}}.

\bibitem{Altmannshofer:2021qrr}
W.~Altmannshofer and P.~Stangl, {\it {New Physics in Rare B Decays after
  Moriond 2021}},  \href{http://arXiv.org/abs/2103.13370}{{\tt 2103.13370}}.

\bibitem{Alguero:2021anc}
M.~Alguer\'o, B.~Capdevila, S.~Descotes-Genon, J.~Matias and M.~Novoa-Brunet,
  {\it {$\boldsymbol{b\to s\ell\ell}$ global fits after Moriond 2021 results}},
   in {\em {55th Rencontres de Moriond on QCD and High Energy Interactions}},
  4, 2021.
\newblock \href{http://arXiv.org/abs/2104.08921}{{\tt 2104.08921}}.

\bibitem{Allanach:2017bta}
B.~C. Allanach, B.~Gripaios and T.~You, {\it {The case for future hadron
  colliders from $B \to K^{(*)} \mu^+ \mu^-$ decays}},  {\em JHEP} {\bf 03}
  (2018) 021 [\href{http://arXiv.org/abs/1710.06363}{{\tt 1710.06363}}].

\bibitem{Allanach:2018odd}
B.~C. Allanach, T.~Corbett, M.~J. Dolan and T.~You, {\it {Hadron collider
  sensitivity to fat flavourful Z$^\prime$s for $ {R}_{K^{\left(\ast \right)}}
  $}},  {\em JHEP} {\bf 03} (2019) 137
  [\href{http://arXiv.org/abs/1810.02166}{{\tt 1810.02166}}].

\bibitem{Allanach:2019mfl}
B.~C. Allanach, J.~M. Butterworth and T.~Corbett, {\it {Collider constraints on
  Z$^\prime$ models for neutral current B-anomalies}},  {\em JHEP} {\bf 08}
  (2019) 106 [\href{http://arXiv.org/abs/1904.10954}{{\tt 1904.10954}}].

\bibitem{Allanach:2018lvl}
B.~C. Allanach and J.~Davighi, {\it {Third family hypercharge model for $
  {R}_{K^{\left(\ast \right)}} $ and aspects of the fermion mass problem}},
  {\em JHEP} {\bf 12} (2018) 075 [\href{http://arXiv.org/abs/1809.01158}{{\tt
  1809.01158}}].

\bibitem{Chung:2021ekz}
Y.~Chung, {\it {A Flavorful Composite Higgs Model : Connecting the B anomalies
  with the hierarchy problem}},  \href{http://arXiv.org/abs/2108.08511}{{\tt
  2108.08511}}.

\bibitem{Altmannshofer:2014cfa}
W.~Altmannshofer, S.~Gori, M.~Pospelov and I.~Yavin, {\it {Quark flavor
  transitions in $L_\mu-L_\tau$ models}},  {\em Phys. Rev. D} {\bf 89} (2014)
  095033 [\href{http://arXiv.org/abs/1403.1269}{{\tt 1403.1269}}].

\bibitem{Crivellin:2015mga}
A.~Crivellin, G.~D'Ambrosio and J.~Heeck, {\it {Explaining
  $h\to\mu^\pm\tau^\mp$, $B\to K^* \mu^+\mu^-$ and $B\to K \mu^+\mu^-/B\to K
  e^+e^-$ in a two-Higgs-doublet model with gauged $L_\mu-L_\tau$}},  {\em
  Phys. Rev. Lett.} {\bf 114} (2015) 151801
  [\href{http://arXiv.org/abs/1501.00993}{{\tt 1501.00993}}].

\bibitem{Crivellin:2015lwa}
A.~Crivellin, G.~D'Ambrosio and J.~Heeck, {\it {Addressing the LHC flavor
  anomalies with horizontal gauge symmetries}},  {\em Phys. Rev. D} {\bf 91}
  (2015), no.~7 075006 [\href{http://arXiv.org/abs/1503.03477}{{\tt
  1503.03477}}].

\bibitem{Crivellin:2015era}
A.~Crivellin, L.~Hofer, J.~Matias, U.~Nierste, S.~Pokorski and J.~Rosiek, {\it
  {Lepton-flavour violating $B$ decays in generic $Z'$ models}},  {\em Phys.
  Rev. D} {\bf 92} (2015), no.~5 054013
  [\href{http://arXiv.org/abs/1504.07928}{{\tt 1504.07928}}].

\bibitem{Altmannshofer:2015mqa}
W.~Altmannshofer and I.~Yavin, {\it {Predictions for lepton flavor universality
  violation in rare B decays in models with gauged $L_\mu - L_\tau$}},  {\em
  Phys. Rev. D} {\bf 92} (2015), no.~7 075022
  [\href{http://arXiv.org/abs/1508.07009}{{\tt 1508.07009}}].

\bibitem{Sierra:2015fma}
D.~Aristizabal~Sierra, F.~Staub and A.~Vicente, {\it {Shedding light on the
  $b\to s$ anomalies with a dark sector}},  {\em Phys. Rev. D} {\bf 92} (2015),
  no.~1 015001 [\href{http://arXiv.org/abs/1503.06077}{{\tt 1503.06077}}].

\bibitem{Celis:2015ara}
A.~Celis, J.~Fuentes-Martin, M.~Jung and H.~Serodio, {\it {Family nonuniversal
  Z' models with protected flavor-changing interactions}},  {\em Phys. Rev. D}
  {\bf 92} (2015), no.~1 015007 [\href{http://arXiv.org/abs/1505.03079}{{\tt
  1505.03079}}].

\bibitem{Greljo:2015mma}
A.~Greljo, G.~Isidori and D.~Marzocca, {\it {On the breaking of Lepton Flavor
  Universality in B decays}},  {\em JHEP} {\bf 07} (2015) 142
  [\href{http://arXiv.org/abs/1506.01705}{{\tt 1506.01705}}].

\bibitem{Falkowski:2015zwa}
A.~Falkowski, M.~Nardecchia and R.~Ziegler, {\it {Lepton Flavor
  Non-Universality in B-meson Decays from a U(2) Flavor Model}},  {\em JHEP}
  {\bf 11} (2015) 173 [\href{http://arXiv.org/abs/1509.01249}{{\tt
  1509.01249}}].

\bibitem{Chiang:2016qov}
C.-W. Chiang, X.-G. He and G.~Valencia, {\it {Z' model for $b\rightarrow{}sll$
  flavor anomalies}},  {\em Phys. Rev. D} {\bf 93} (2016), no.~7 074003
  [\href{http://arXiv.org/abs/1601.07328}{{\tt 1601.07328}}].

\bibitem{Boucenna:2016wpr}
S.~M. Boucenna, A.~Celis, J.~Fuentes-Martin, A.~Vicente and J.~Virto, {\it
  {Non-abelian gauge extensions for B-decay anomalies}},  {\em Phys. Lett. B}
  {\bf 760} (2016) 214--219 [\href{http://arXiv.org/abs/1604.03088}{{\tt
  1604.03088}}].

\bibitem{Boucenna:2016qad}
S.~M. Boucenna, A.~Celis, J.~Fuentes-Martin, A.~Vicente and J.~Virto, {\it
  {Phenomenology of an $SU(2) \times SU(2) \times U(1)$ model with
  lepton-flavour non-universality}},  {\em JHEP} {\bf 12} (2016) 059
  [\href{http://arXiv.org/abs/1608.01349}{{\tt 1608.01349}}].

\bibitem{Ko:2017lzd}
P.~Ko, Y.~Omura, Y.~Shigekami and C.~Yu, {\it {LHCb anomaly and B physics in
  flavored Z' models with flavored Higgs doublets}},  {\em Phys. Rev. D} {\bf
  95} (2017), no.~11 115040 [\href{http://arXiv.org/abs/1702.08666}{{\tt
  1702.08666}}].

\bibitem{Alonso:2017bff}
R.~Alonso, P.~Cox, C.~Han and T.~T. Yanagida, {\it {Anomaly-free local
  horizontal symmetry and anomaly-full rare B-decays}},  {\em Phys. Rev. D}
  {\bf 96} (2017), no.~7 071701 [\href{http://arXiv.org/abs/1704.08158}{{\tt
  1704.08158}}].

\bibitem{Tang:2017gkz}
Y.~Tang and Y.-L. Wu, {\it {Flavor non-universal gauge interactions and
  anomalies in B-meson decays}},  {\em Chin. Phys. C} {\bf 42} (2018), no.~3
  033104 [\href{http://arXiv.org/abs/1705.05643}{{\tt 1705.05643}}]. [Erratum:
  Chin.Phys.C 44, 069101 (2020)].

\bibitem{Bhatia:2017tgo}
D.~Bhatia, S.~Chakraborty and A.~Dighe, {\it {Neutrino mixing and $R_K$ anomaly
  in U(1)$_X$ models: a bottom-up approach}},  {\em JHEP} {\bf 03} (2017) 117
  [\href{http://arXiv.org/abs/1701.05825}{{\tt 1701.05825}}].

\bibitem{Fuyuto:2017sys}
K.~Fuyuto, H.-L. Li and J.-H. Yu, {\it {Implications of hidden gauged $U(1)$
  model for $B$ anomalies}},  {\em Phys. Rev. D} {\bf 97} (2018), no.~11 115003
  [\href{http://arXiv.org/abs/1712.06736}{{\tt 1712.06736}}].

\bibitem{Bian:2017xzg}
L.~Bian, H.~M. Lee and C.~B. Park, {\it {$B$-meson anomalies and Higgs physics
  in flavored $U(1)'$ model}},  {\em Eur. Phys. J. C} {\bf 78} (2018), no.~4
  306 [\href{http://arXiv.org/abs/1711.08930}{{\tt 1711.08930}}].

\bibitem{Alonso:2017uky}
R.~Alonso, P.~Cox, C.~Han and T.~T. Yanagida, {\it {Flavoured $B-L$ local
  symmetry and anomalous rare $B$ decays}},  {\em Phys. Lett. B} {\bf 774}
  (2017) 643--648 [\href{http://arXiv.org/abs/1705.03858}{{\tt 1705.03858}}].

\bibitem{Bonilla:2017lsq}
C.~Bonilla, T.~Modak, R.~Srivastava and J.~W.~F. Valle, {\it
  {$U(1)_{B_3-3L_\mu}$ gauge symmetry as a simple description of $b\to s$
  anomalies}},  {\em Phys. Rev. D} {\bf 98} (2018), no.~9 095002
  [\href{http://arXiv.org/abs/1705.00915}{{\tt 1705.00915}}].

\bibitem{King:2018fcg}
S.~F. King, {\it {$ {R}_{K^{\left(*\right)}} $ and the origin of Yukawa
  couplings}},  {\em JHEP} {\bf 09} (2018) 069
  [\href{http://arXiv.org/abs/1806.06780}{{\tt 1806.06780}}].

\bibitem{Duan:2018akc}
G.~H. Duan, X.~Fan, M.~Frank, C.~Han and J.~M. Yang, {\it {A minimal
  $U(1)^\prime$ extension of MSSM in light of the B decay anomaly}},  {\em
  Phys. Lett. B} {\bf 789} (2019) 54--58
  [\href{http://arXiv.org/abs/1808.04116}{{\tt 1808.04116}}].

\bibitem{Kang:2019vng}
Z.~Kang and Y.~Shigekami, {\it {$(g-2)_{\mu}$ versus flavor changing neutral
  current induced by the light $(B-L)_{\mu\tau}$ boson}},  {\em JHEP} {\bf 11}
  (2019) 049 [\href{http://arXiv.org/abs/1905.11018}{{\tt 1905.11018}}].

\bibitem{Calibbi:2019lvs}
L.~Calibbi, A.~Crivellin, F.~Kirk, C.~A. Manzari and L.~Vernazza, {\it
  {$Z^\prime$ models with less-minimal flavour violation}},  {\em Phys. Rev. D}
  {\bf 101} (2020), no.~9 095003 [\href{http://arXiv.org/abs/1910.00014}{{\tt
  1910.00014}}].

\bibitem{Altmannshofer:2019xda}
W.~Altmannshofer, J.~Davighi and M.~Nardecchia, {\it {Gauging the accidental
  symmetries of the standard model, and implications for the flavor
  anomalies}},  {\em Phys. Rev. D} {\bf 101} (2020), no.~1 015004
  [\href{http://arXiv.org/abs/1909.02021}{{\tt 1909.02021}}].

\bibitem{Capdevila:2020rrl}
B.~Capdevila, A.~Crivellin, C.~A. Manzari and M.~Montull, {\it {Explaining
  $b\to s\ell^+\ell^-$ and the Cabibbo angle anomaly with a vector triplet}},
  {\em Phys. Rev. D} {\bf 103} (2021), no.~1 015032
  [\href{http://arXiv.org/abs/2005.13542}{{\tt 2005.13542}}].

\bibitem{Davighi:2020qqa}
J.~Davighi, M.~Kirk and M.~Nardecchia, {\it {Anomalies and accidental
  symmetries: charging the scalar leptoquark under L$_{\mu}$ \ensuremath{-}
  L$_{\tau}$}},  {\em JHEP} {\bf 12} (2020) 111
  [\href{http://arXiv.org/abs/2007.15016}{{\tt 2007.15016}}].

\bibitem{Allanach:2020kss}
B.~C. Allanach, {\it {$U(1)_{B_3-L_2}$ explanation of the neutral current
  $B$\ensuremath{-}anomalies}},  {\em Eur. Phys. J. C} {\bf 81} (2021), no.~1
  56 [\href{http://arXiv.org/abs/2009.02197}{{\tt 2009.02197}}]. [Erratum:
  Eur.Phys.J.C 81, 321 (2021)].

\bibitem{Borah:2020swo}
D.~Borah, L.~Mukherjee and S.~Nandi, {\it {Low scale U(1)$_{X}$ gauge symmetry
  as an origin of dark matter, neutrino mass and flavour anomalies}},  {\em
  JHEP} {\bf 12} (2020) 052 [\href{http://arXiv.org/abs/2007.13778}{{\tt
  2007.13778}}].

\bibitem{Bednyakov:2021fof}
A.~Bednyakov and A.~Mukhaeva, {\it {Flavour Anomalies in a $U(1)$ SUSY
  Extension of the SM}},  {\em Symmetry} {\bf 13} (2021), no.~2 191.

\bibitem{Davighi:2021oel}
J.~Davighi, {\it {Anomalous $Z^\prime$ bosons for anomalous $B$ decays}},
  \href{http://arXiv.org/abs/2105.06918}{{\tt 2105.06918}}.

\bibitem{Greljo:2021npi}
A.~Greljo, Y.~Soreq, P.~Stangl, A.~E. Thomsen and J.~Zupan, {\it {Muonic Force
  Behind Flavor Anomalies}},  \href{http://arXiv.org/abs/2107.07518}{{\tt
  2107.07518}}.

\bibitem{Wang:2021uqz}
X.~Wang, {\it {Muon $(g-2)$ and Flavor Puzzles in the $U(1)^{}_{X}$-gauged
  Leptoquark Model}},  \href{http://arXiv.org/abs/2108.01279}{{\tt
  2108.01279}}.

\bibitem{Bhatia:2021eco}
D.~Bhatia, N.~Desai and A.~Dighe, {\it {Frugal $U(1)_X$ models with non-minimal
  flavor violation for $b \to s \ell \ell$ anomalies and neutrino mixing}},
  \href{http://arXiv.org/abs/2109.07093}{{\tt 2109.07093}}.

\bibitem{Allanach:2019iiy}
B.~C. Allanach and J.~Davighi, {\it {Naturalising the third family hypercharge
  model for neutral current $B$-anomalies}},  {\em Eur. Phys. J. C} {\bf 79}
  (2019), no.~11 908 [\href{http://arXiv.org/abs/1905.10327}{{\tt
  1905.10327}}].

\bibitem{Allanach:2021kzj}
B.~C. Allanach, J.~E. Camargo-Molina and J.~Davighi, {\it {Global fits of third
  family hypercharge models to neutral current B-anomalies and electroweak
  precision observables}},  {\em Eur. Phys. J. C} {\bf 81} (2021), no.~8 721
  [\href{http://arXiv.org/abs/2103.12056}{{\tt 2103.12056}}].

\bibitem{Butterworth:2016sqg}
J.~M. Butterworth, D.~Grellscheid, M.~Kr\"amer, B.~Sarrazin and D.~Yallup, {\it
  {Constraining new physics with collider measurements of Standard Model
  signatures}},  {\em JHEP} {\bf 03} (2017) 078
  [\href{http://arXiv.org/abs/1606.05296}{{\tt 1606.05296}}].

\bibitem{Buckley:2021neu}
A.~Buckley {\em et.~al.}, {\it {Testing new physics models with global
  comparisons to collider measurements: the Contur toolkit}},  {\em SciPost
  Phys. Core} {\bf 4} (2021) 013 [\href{http://arXiv.org/abs/2102.04377}{{\tt
  2102.04377}}].

\bibitem{ParticleDataGroup:2020ssz}
{\bf Particle Data Group} Collaboration, P.~A. Zyla {\em et.~al.}, {\it {Review
  of Particle Physics}},  {\em PTEP} {\bf 2020} (2020), no.~8 083C01.

\bibitem{Degrande:2011ua}
C.~Degrande, C.~Duhr, B.~Fuks, D.~Grellscheid, O.~Mattelaer and T.~Reiter, {\it
  {UFO - The Universal FeynRules Output}},  {\em Comput. Phys. Commun.} {\bf
  183} (2012) 1201--1214 [\href{http://arXiv.org/abs/1108.2040}{{\tt
  1108.2040}}].

\bibitem{Alloul:2013bka}
A.~Alloul, N.~D. Christensen, C.~Degrande, C.~Duhr and B.~Fuks, {\it {FeynRules
  2.0 - A complete toolbox for tree-level phenomenology}},  {\em Comput. Phys.
  Commun.} {\bf 185} (2014) 2250--2300
  [\href{http://arXiv.org/abs/1310.1921}{{\tt 1310.1921}}].

\bibitem{Bellm:2015jjp}
J.~Bellm {\em et.~al.}, {\it {Herwig 7.0/Herwig++ 3.0 release note}},  {\em
  Eur. Phys. J.} {\bf C76} (2016), no.~4 196
  [\href{http://arXiv.org/abs/1512.01178}{{\tt 1512.01178}}].
%%CITATION = ARXIV:1512.01178;%%

\bibitem{Bahr:2008pv}
M.~Bahr {\em et.~al.}, {\it {Herwig++ Physics and Manual}},  {\em Eur. Phys.
  J.} {\bf C58} (2008) 639--707 [\href{http://arXiv.org/abs/0803.0883}{{\tt
  0803.0883}}].
%%CITATION = ARXIV:0803.0883;%%

\bibitem{Bierlich:2019rhm}
C.~Bierlich, A.~Buckley, J.~M. Butterworth, L.~Corpe, D.~Grellscheid,
  C.~Gutschow, P.~Karczmarczyk, J.~Klein, L.~Lonnblad, C.~S. Pollard, H.~Schulz
  and F.~Siegert, {\it {Robust Independent Validation of Experiment and Theory:
  Rivet version 3}},  {\em SciPost Phys.} {\bf 8} (2020) 026
  [\href{http://arXiv.org/abs/1912.05451}{{\tt 1912.05451}}].
%%CITATION = ARXIV:1912.05451;%%

\bibitem{Maguire:2017ypu}
E.~Maguire, L.~Heinrich and G.~Watt, {\it {HEPData: a repository for high
  energy physics data}},  {\em J. Phys. Conf. Ser.} {\bf 898} (2017), no.~10
  102006 [\href{http://arXiv.org/abs/1704.05473}{{\tt 1704.05473}}].
%%CITATION = ARXIV:1704.05473;%%

\bibitem{Dulat:2015mca}
S.~Dulat, T.-J. Hou, J.~Gao, M.~Guzzi, J.~Huston, P.~Nadolsky, J.~Pumplin,
  C.~Schmidt, D.~Stump and C.~P. Yuan, {\it {New parton distribution functions
  from a global analysis of quantum chromodynamics}},  {\em Phys. Rev. D} {\bf
  93} (2016), no.~3 033006 [\href{http://arXiv.org/abs/1506.07443}{{\tt
  1506.07443}}].

\bibitem{CMS:2018mdl}
{\bf CMS} Collaboration, A.~M. Sirunyan {\em et.~al.}, {\it {Measurement of the
  differential Drell-Yan cross section in proton-proton collisions at $
  \sqrt{\mathrm{s}} $ = 13 TeV}},  {\em JHEP} {\bf 12} (2019) 059
  [\href{http://arXiv.org/abs/1812.10529}{{\tt 1812.10529}}].

\bibitem{ATLAS:2019gey}
{\bf ATLAS} Collaboration, G.~Aad {\em et.~al.}, {\it {Measurement of the
  $Z(\rightarrow\ell^+\ell^-)\gamma$ production cross-section in $pp$
  collisions at $\sqrt{s} =13$ TeV with the ATLAS detector}},  {\em JHEP} {\bf
  03} (2020) 054 [\href{http://arXiv.org/abs/1911.04813}{{\tt 1911.04813}}].

\bibitem{ATLAS:2016qjc}
{\bf ATLAS} Collaboration, G.~Aad {\em et.~al.}, {\it {Measurements of
  $Z\gamma$ and $Z\gamma\gamma$ production in $pp$ collisions at $\sqrt{s}=$ 8
  TeV with the ATLAS detector}},  {\em Phys. Rev. D} {\bf 93} (2016), no.~11
  112002 [\href{http://arXiv.org/abs/1604.05232}{{\tt 1604.05232}}].

\bibitem{ATLAS:2013way}
{\bf ATLAS} Collaboration, G.~Aad {\em et.~al.}, {\it {Measurements of $W
  \gamma$ and $Z \gamma$ production in $pp$ collisions at $\sqrt{s}$=7 TeV with
  the ATLAS detector at the LHC}},  {\em Phys. Rev. D} {\bf 87} (2013), no.~11
  112003 [\href{http://arXiv.org/abs/1302.1283}{{\tt 1302.1283}}]. [Erratum:
  Phys.Rev.D 91, 119901 (2015)].

\bibitem{ATLAS:2017sag}
{\bf ATLAS} Collaboration, M.~Aaboud {\em et.~al.}, {\it {Measurements of the
  production cross section of a $Z$ boson in association with jets in pp
  collisions at $\sqrt{s} = 13$ TeV with the ATLAS detector}},  {\em Eur. Phys.
  J. C} {\bf 77} (2017), no.~6 361 [\href{http://arXiv.org/abs/1702.05725}{{\tt
  1702.05725}}].

\bibitem{CMS:2018mdf}
{\bf CMS} Collaboration, A.~M. Sirunyan {\em et.~al.}, {\it {Measurement of
  differential cross sections for Z boson production in association with jets
  in proton-proton collisions at $\sqrt{s} =$ 13 TeV}},  {\em Eur. Phys. J. C}
  {\bf 78} (2018), no.~11 965 [\href{http://arXiv.org/abs/1804.05252}{{\tt
  1804.05252}}].

\bibitem{LHCb:2016nhs}
{\bf LHCb} Collaboration, R.~Aaij {\em et.~al.}, {\it {Measurement of forward
  $W$ and $Z$ boson production in association with jets in proton-proton
  collisions at $\sqrt{s}=8$ TeV}},  {\em JHEP} {\bf 05} (2016) 131
  [\href{http://arXiv.org/abs/1605.00951}{{\tt 1605.00951}}].

\bibitem{CMS:2018dxg}
{\bf CMS} Collaboration, A.~M. Sirunyan {\em et.~al.}, {\it {Measurement of
  associated production of a W boson and a charm quark in proton-proton
  collisions at $\sqrt{s} =$ 13 TeV}},  {\em Eur. Phys. J. C} {\bf 79} (2019),
  no.~3 269 [\href{http://arXiv.org/abs/1811.10021}{{\tt 1811.10021}}].

\bibitem{ATLAS:2019hxz}
{\bf ATLAS} Collaboration, G.~Aad {\em et.~al.}, {\it {Measurements of
  top-quark pair differential and double-differential cross-sections in the
  $\ell$+jets channel with $pp$ collisions at $\sqrt{s}=13$ TeV using the ATLAS
  detector}},  {\em Eur. Phys. J. C} {\bf 79} (2019), no.~12 1028
  [\href{http://arXiv.org/abs/1908.07305}{{\tt 1908.07305}}]. [Erratum:
  Eur.Phys.J.C 80, 1092 (2020)].

\bibitem{ATLAS:2014ape}
{\bf ATLAS} Collaboration, G.~Aad {\em et.~al.}, {\it {Measurement of the
  low-mass Drell-Yan differential cross section at $\sqrt{s}$ = 7 TeV using the
  ATLAS detector}},  {\em JHEP} {\bf 06} (2014) 112
  [\href{http://arXiv.org/abs/1404.1212}{{\tt 1404.1212}}].

\bibitem{ATLAS:2021kog}
{\bf ATLAS} Collaboration, G.~Aad {\em et.~al.}, {\it {Measurements of
  differential cross-sections in four-lepton events in 13 TeV proton-proton
  collisions with the ATLAS detector}},  {\em JHEP} {\bf 07} (2021) 005
  [\href{http://arXiv.org/abs/2103.01918}{{\tt 2103.01918}}].

\bibitem{Abi:2021gix}
{\bf Muon $g-2$ Collboration} Collaboration, B.~Abi {\em et.~al.}, {\it
  {Measurement of the Positive Muon Anomalous Magnetic Moment to 0.46~ppm}},
  {\em Phys. Rev. Lett.} {\bf 126} (2021) 141801
  [\href{http://arXiv.org/abs/2104.03281}{{\tt 2104.03281}}].

\bibitem{Jegerlehner:2009ry}
F.~Jegerlehner and A.~Nyffeler, {\it {The Muon g-2}},  {\em Phys. Rept.} {\bf
  477} (2009) 1--110 [\href{http://arXiv.org/abs/0902.3360}{{\tt 0902.3360}}].

\bibitem{ATLAS:2017mpa}
{\bf ATLAS} Collaboration, {\it {Measurement of the $\tau$ lepton
  reconstruction and identification performance in the ATLAS experiment using
  $pp$ collisions at $\sqrt{s}=13~{ TeV}$}},  Tech. Rep. ATLAS-CONF-2017-029,
  May, 2017.

\bibitem{CMS:2018rym}
{\bf CMS} Collaboration, A.~M. Sirunyan {\em et.~al.}, {\it {Performance of the
  CMS muon detector and muon reconstruction with proton-proton collisions at
  $\sqrt{s}=$ 13 TeV}},  {\em JINST} {\bf 13} (2018), no.~06 P06015
  [\href{http://arXiv.org/abs/1804.04528}{{\tt 1804.04528}}].

\bibitem{ATLAS:2020auj}
{\bf ATLAS} Collaboration, G.~Aad {\em et.~al.}, {\it {Muon reconstruction and
  identification efficiency in ATLAS using the full Run 2 $pp$ collision data
  set at $\sqrt{s}=13$ TeV}},  {\em Eur. Phys. J. C} {\bf 81} (2021), no.~7 578
  [\href{http://arXiv.org/abs/2012.00578}{{\tt 2012.00578}}].

\end{thebibliography}\endgroup

\end{document}